\documentclass[a4paper,11pt]{article}
\usepackage{pos}
\usepackage{float}

\usepackage{amsmath}
\usepackage{physics}
\usepackage{hyperref}

\title{Resolving the critical bubble in $\rm{SU}(8)$ deconfinement transition}

\author{Kari Rummukainen}
\author*{Riikka Seppä}
\author{David J. Weir}

\affiliation{Department of Physics and Helsinki Institute of Physics, P.O.~Box 64, FI-00014 University 
	of Helsinki, Finland}

\emailAdd{kari.rummukainen@helsinki.fi}
\emailAdd{riikka.seppa@helsinki.fi}
\emailAdd{david.weir@helsinki.fi}

\abstract{

Strongly coupled confining models with a first order phase transition present an interesting DM candidate.
Near the critical temperature these models are strongly non-perturbative, and the critical bubble nucleation rate has so far only been estimated via approximative methods. As a model of a strong first order deconfinement transition, we simulate 4D SU(8) pure gauge model with multicanonical Monte Carlo. We demonstrate that resolving the critical bubble is possible in a strongly coupled model.
 We calculate the free energy of a critical bubble, which gives a rough upper limit for the nucleation rate. 
For the parameter points we investigated, the thin-wall approximation for the critial bubble free energy is off by a factor of 2, overestimating the rate at least by a factor of $e^{10}$ .

}

\FullConference{The 41st International Symposium on Lattice Field Theory (LATTICE2024)\\
 28 July - 3 August 2024\\
Liverpool, UK\\}

\begin{document}

\renewcommand{\logo}{\relax}

\maketitle

\section{Introduction}

In recent years, many confining dark matter models with a first order phase transition have been proposed, see e.g. Refs.~\cite{Huang:2020crf, LatticeStrongDynamics:2020jwi, Bruno:2024dha}. As strongly coupled confining models are non-perturbative near the critical temperature, the nucleation rate cannot be computed perturbatively. With lattice inputs, it can be estimated semi-classically with the thin-wall approximation or effective models \cite{Huang:2020crf}. In weakly coupled theories, the whole rate could be computed purely from the lattice with the use of real-time simulations \cite{Moore:2000jw}. Even without real-time simulations an estimate of the nucleation rate can be computed by measuring the probability of a critical bubble configuration, which gives the free energy of the critical bubble.

In this proceedings, we explore the deconfinement transition of the SU(8) pure gauge model, using finite temperature Monte Carlo simulations, with the goal of measuring the free energy of a critical bubble configuration. In practice this means measuring the probability of a critical bubble configuration with respect to the metastable, in this case confining, phase. The mixed-phase bubble configurations are extremely suppressed compared to the bulk phase configurations, which prompts us to employ the multicanonical algorithm. 

We use SU(8) pure gauge theory as a prototype model for a strongly coupled confining model. We investigate the transition from the confining to the deconfining phase, i.e. from a `cold' state to a `hot' state. In the context of early universe phase transitions, the opposite direction would be a more physically motivated scenario. Still, our study serves as a proof of principle that the nucleation rate of a strongly coupled model can be estimated directly from the lattice. 

Our choice of number of colors $N_c = 8$, while making the simulations computationally more demanding, is due to the transition being stronger and the correlation length shorter than for $N_c < 8$ \citep{Lucini:2005vg}. This allows larger superheating and -cooling, in turn making the critical bubbles smaller and easier to contain on a finite lattice.
 However, we still run into troublesome effects from having to use large lattices. We show that these effects can be lessened by smearing and by choosing a more suitable order parameter. We obtain a preliminary result for the nucleation rate at two different levels of superheating.

\section{Lattice setup}

We study the transition on a periodic $L_s^3 L_t$ lattice with lattice spacing $a$, so that $a N_s = L_s$, and $aN_t = L_t$.
The action is the standard plaquette action
\begin{align}
   S = \beta \sum_{x, \mu > \nu} \left[  1 - \frac{1}{8}\Re \Tr U_{\mu\nu}(x) \right],
\end{align}
where $U_{\mu\nu}(x)$ is the Wilson plaquette at $x$, with $U_{\mu}(x) \in \mathrm{SU}(8)$.
The Euclidean time direction is related to the temperature through $T = 1/(N_t a(\beta))$.
If $N_t$ is held constant, the temperature is a function of $\beta$ only. The critical value of $\beta$ is well known for this model \cite{Lucini:2012wq}. We use $\beta_c = 44.562$ \cite{surfacetension}.  

We work with lattices of size $N_s = 60$, $N_t = 6$, and study the system at couplings $\beta_1 = 44.712$ and $\beta_2 = 44.742$. These correspond to $\Delta \beta = \beta - \beta_c = 0.15$ and $0.18$, respectively. We use a standard mix of heatbath and overrelaxation updates, with 5 to 6 overrelaxation updates for each heatbath,  and a multicanonical accept-reject step after updating any time-direction links.

\subsection{Critical bubble free energy} \label{sect:nucleratecalc}

We use the nonperturbative method introduced in Refs.~\cite{Moore:2000jw, Moore:2001vf}, more recently used e.g. in Refs.~\cite{Gould:2022ran,Gould:2024chm} to calculate the critical bubble free energy. The bubble configurations are extremely suppressed with respect to the bulk phase configurations. We employ multicanonical Monte Carlo \cite{Ferrenberg:1988yz, Ferrenberg:1989ui, Berg:1992qua}, with automatic tuning as in \cite{Moore:2000jw}. Instead of sampling with $p \propto \exp(-S)$ as in normal Monte Carlo, in multicanonical method we sample with $p_{\mathrm{muca}} \propto \exp[-S + W(\mathcal{O})]$, where $\mathcal{O}$ is our order parameter, and $W$ a weight function iteratively constructed to make the probability distribution of $\mathcal{O}$ obtained from the simulation approximately constant.

By reweighting $P_{\mathrm{muca}}(\mathcal{O})$, the measured distribution of configurations with an order parameter value $\mathcal{O}$, with the weight function $W(\mathcal{O})$, we arrive back to the canonical distribution $P(\mathcal{O})$,
\begin{align}
P(\mathcal{O}) \propto e^{-W(\mathcal{O})} P_{\mathrm{muca}}(\mathcal{O}). \label{eq:canon_distro}
\end{align}

From the reweighted probability density $P(\mathcal{O})$, in practice a histogram, the free energy of the critical bubble can be measured as \cite{Moore:2001vf}
\begin{align}
\dfrac{F_C}{T} \approx \dfrac{F(\mathcal{O}_C)}{T} - \dfrac{F(\mathcal{O}_\mathrm{conf})}{T} = -\log \dfrac{P(\mathcal{O}_C)}{P(\mathcal{O}_\mathrm{conf})},
\end{align}
where $\mathcal{O}_C$ is the value of the order parameter at the critical bubble, and $\mathcal{O}_\mathrm{conf}$ the value at the confining phase peak. The critical bubble is the most suppressed configuration of the transition, so in essence $P(\mathcal{O}_C)$ corresponds to the minimum value of the histogram, and $P(\mathcal{O}_\mathrm{conf})$ to the maximum value on the confinement phase peak.\footnote{It would be more proper to integrate over the histogram values, but this way we get a dimensionless quantity.}
 
\subsection{Finite size effects}

\begin{figure}[h]
   \centering \includegraphics[width=0.75\textwidth]{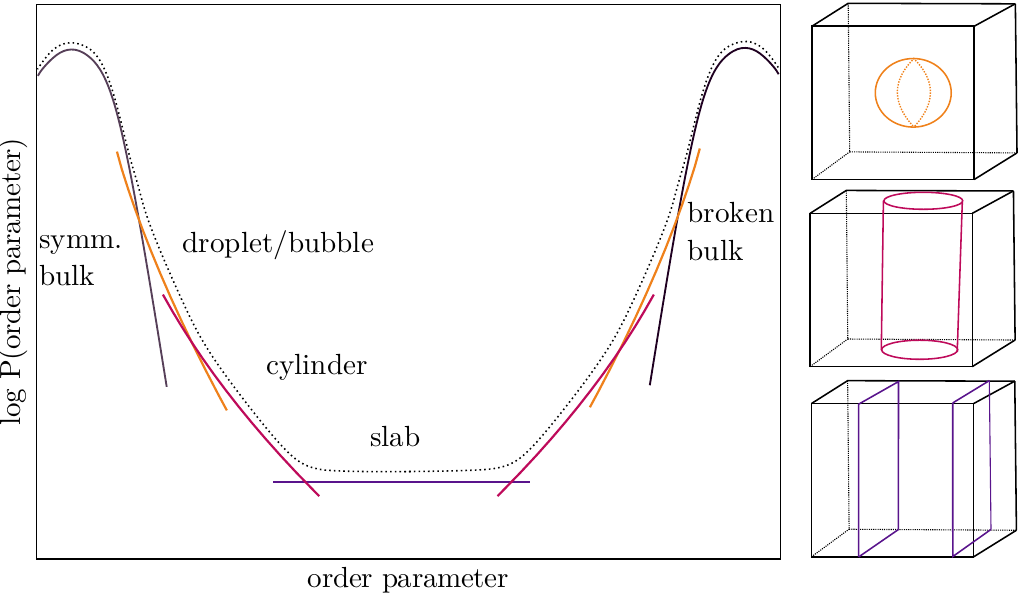} 
   \caption{Schematic of different topological regimes/branches predicted by the thin-wall approximation in a box with periodic boundaries, at the critical temperature. Above the critical temperature the broken phase peak grows larger than the symmetric peak. For large lattices, the bulk phase fluctuations can be sufficiently broad as to cover the curve corresponding to the droplet branch. Based on Fig. 1 of Ref.~\cite{Moore:2001vf}.}
   \label{fig:thinw}
\end{figure}

In above discussion we have implicitly assumed that there exists some value of the order parameter where we have a critical bubble configuration on the lattice, i.e. a bubble configuration for which the order parameter is a minimum of the probability density. For a finite sized periodic lattice this does not necessarily hold. If the bubble is too large compared to the lattice size, it will in practice settle into a cylinder or a slab. By estimating the topological regimes with the thin-wall approximation, illustrated in Fig.~\ref{fig:thinw}, a mixed-phase configuration where the nucleating stable phase occupies up to $4\pi /81$ of the volume should generally have a spherical topology \cite{Moore:2000jw}.

Moving above the critical temperature, i.e.~superheating the system, will make the critical bubble smaller. However, we want to obtain the nucleation rate as close to the critical temperature as possible, so when trying to `fit the bubble on the lattice', we prefer larger volumes over larger superheating.   

Increasing the volume introduces another problem. The fluctuations of the order parameter grow as square root of the volume, so the order parameter becomes worse at `recognizing' bubble configurations. With a `bad' order parameter, the bubble regime will eventually be swallowed by the bulk phase fluctuations \cite{Moore:2001vf}. This leaves us unable to resolve the critical bubble configurations, but the problem can potentially be alleviated by choosing a better order parameter.

\subsection{Order parameter}	

The Polyakov loop at spatial lattice site $\vec{x}$ is defined as
\begin{align}
l_p(\vec{x}) = \Tr \prod^{N_t -1}_{t = 0} U_4(\vec{x}, t).\label{eq:Polyakov_loop}
\end{align}	  
The conventional order parameter when considering the deconfinement-confinement transition in a pure gauge model is the volume average of the Polyakov loop,

\begin{align}
\langle l_p \rangle = \left\lvert \dfrac{1}{N_s^3} \sum_{\vec{x}}  l_p(\vec{x})\right\rvert. \label{eq:old_order_parameter}
\end{align}	
Note that for $N = 8$, the Polyakov loop is complex, which is why we take the absolute value.

If, for the volume and $\Delta\beta$ we want to investigate, $\langle l_p \rangle$ is used as the order parameter, the Polyakov loop fluctuates so much that the mean value of the order parameter in the presence of the bubble is well within the fluctuations of the order parameter in the bulk phase. Thus, it is impossible to use $\langle l_p \rangle$ to distinguish the existence of a bubble. To alleviate this problem, we investigate modified order parameters which aim to minimize the impact of fluctuations. The new order parameter is used both to iteratively generate the weight function $W$ and to measure the distribution of configurations.

We first smear the Polyakov loop field defined by Eq.~(\ref{eq:Polyakov_loop}), which gets rid of UV fluctuations. In this work, we perform either 24 or 48 smearing steps, with one smearing step constituting to taking the weighted average over nearest neighbours for each spatial lattice site,
\begin{align}
l^{n+1}_{p,s}(\vec{x}) = \dfrac{1}{4} \left( l^n_{p,s}(\vec{x}) + \dfrac{1}{2} \sum_{\hat{i}} l^n_{p,s}(\vec{x} + \hat{i}) \right),
\end{align}
where $l^0_{p,s} = l_p$. With the resulting smeared Polyakov loops $l_s(\vec{x})$ we can construct 
\begin{align}
\langle l_{s}^2 \rangle = \dfrac{1}{N_s^3} \sum_{\vec{x}}  |l_s(\vec{x})|^2.
\end{align} 

The smearing and the squaring already make the bulk phase fluctuation peak less wide, but we can go further and construct two additional candidates for an improved order parameter. First of these is
\begin{align}
\langle l_{\theta} \rangle =  \dfrac{1}{N_s^3} \left\lbrace  \sum_{\vec{x}}  |l_s(\vec{x})|^2 - 2A \sum_{\vec{x}} | l_s(\vec{x}) | \right\rbrace, \label{eq:theta_order_parameter}
\end{align}
where the constant $A$ is chosen to correspond to the position of the bulk phase peak of $\langle  |l_{s}| \rangle$. A similar order parameter was recently applied in Ref.~\cite{Gould:2024chm}.

The second (potentially) improved order parameter we construct is related to the susceptibility of the Polyakov loop value over a given lattice configuration,
\begin{align}
\langle l_{\sigma} \rangle =  \dfrac{1}{N_s^3}  \sum_{\vec{x}}  |l_s(\vec{x})|^2 - \left( \dfrac{1}{N_s^3}\sum_{\vec{x}} | l_s(\vec{x}) | \right) ^2 .
\end{align}
This `order' parameter is not strictly speaking a `true' order parameter, since it cannot differentiate between the confined and deconfined phases, but rather it differentiates between bulk phase and mixed-phase configurations. However, without any loss of generality in the bubble probability measurement, by a suitable choice of weight function we can restrict the range of our simulation to only the confined (metastable) bulk peak and mixed-phase configurations up to bubbles slightly larger than the critical bubble. Then it does not matter that $\langle l_{\sigma} \rangle$ does not differentiate the bulk phases.

\section{Results}

\begin{figure}[h]
   \centering 
   \hspace{2mm}
   \includegraphics[width=0.31\textwidth]{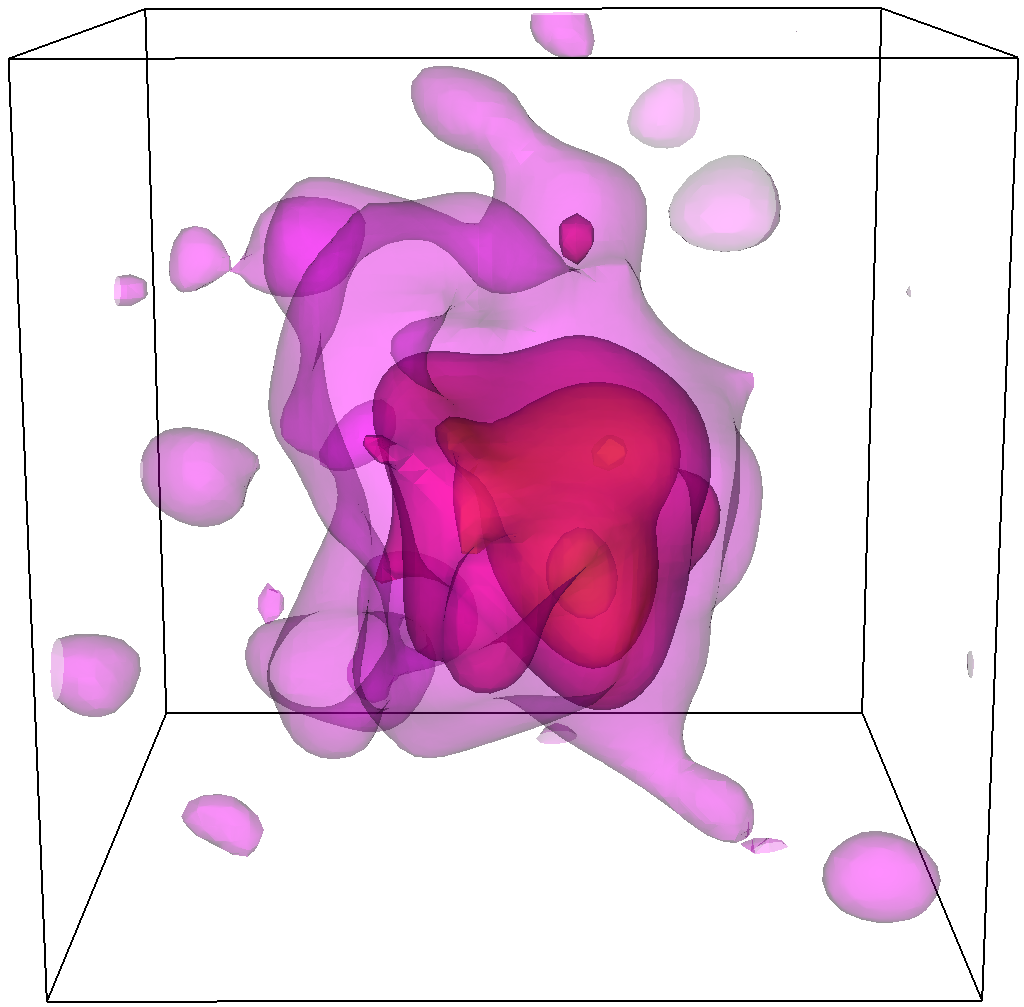}
   \hspace{2mm}
   \includegraphics[width=0.31\textwidth]{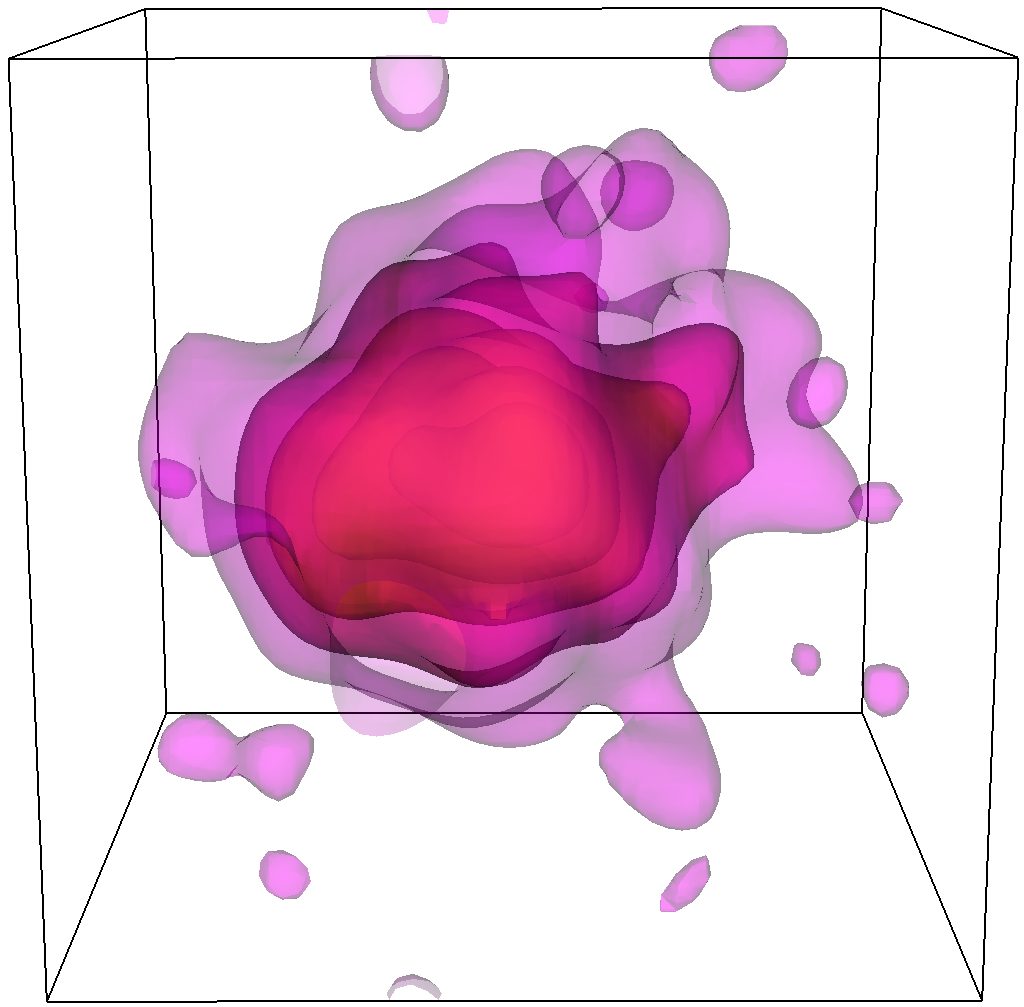}
   \hspace{2mm}
   \includegraphics[width=0.31\textwidth]{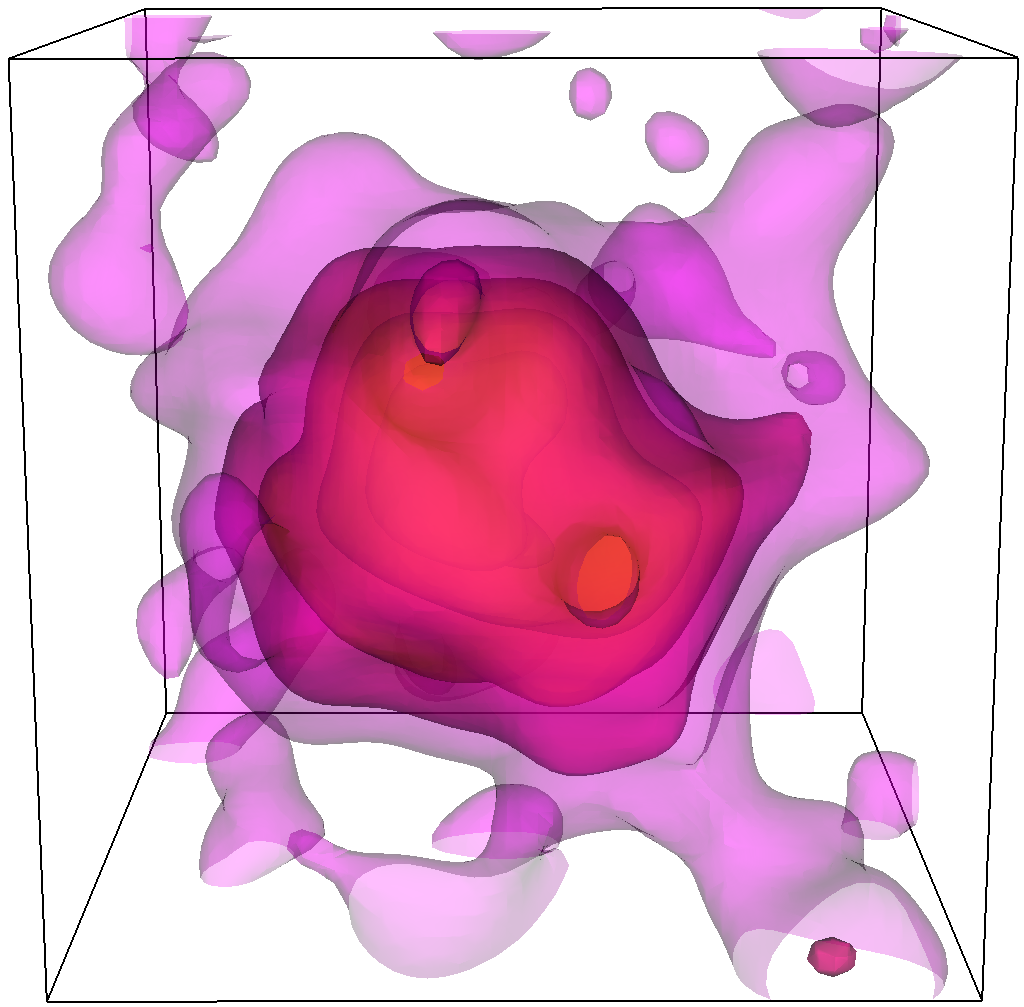}
   \caption{Snapshots of bubble configurations at $N_s = 60, N_t = 6, \beta_2 = 44.742$.  Isosurfaces of the \textit{smeared} Polyakov loop $|l_s|$ at each spatial lattice site are shown, with 48 steps of smearing. From left to right: configurations 1, 2 and 3 of Fig. \ref{fig:lp-ltheta}.}
   \label{fig:vizs-three}
\end{figure}

\begin{figure}[h]
   \centering \includegraphics[width=0.95\textwidth]{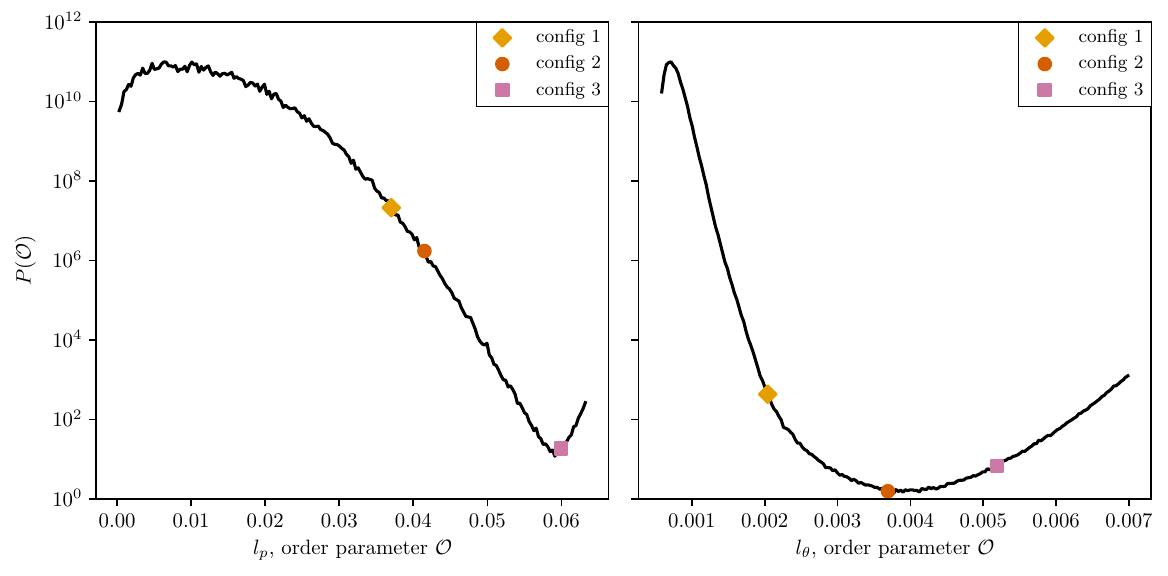}
   \caption{A comparison of probability distributions obtained by using different order parameters, both for the weight function iteration and measurements. Obtained at $\beta_2 = 44.742$ with  \textbf{Left:} $\mathcal{O} = \langle l_p \rangle$ (non-smeared). \textbf{Right:} $\mathcal{O} = \langle l_{\theta} \rangle$ (smeared with 48 steps, A = 0.05). Note that the order parameter values are not directly comparable. Order parameter values corresponding to those calculated from configurations 1, 2, 3 of Fig. \ref{fig:vizs-three} are shown with markers, from left to right.} 
  \label{fig:lp-ltheta}
\end{figure}

We first need to confirm that the minimum of the measured probability distribution $P(\mathcal{O})$ corresponds to a bubble configuration. Based purely on the thin-wall approximation, at $\beta_2 = 44.742$ the minimum should be on the bubble branch, but at $\beta_1 = 44.712$ the minimum is close to the intersection of the bubble and the cylinder branches, so we expect to see more finite size effects. 

Based on the thin-wall estimate of the location of the minimum, and how wide we observed the bulk peak to be for the non-smeared order parameter $\langle l_p \rangle$, we expect that for both $\beta$ the bulk phase fluctuations almost completely cover up the bubble branch. This can further be confirmed by inspecting the configurations we know to be bubble configurations, and checking where they fall on the measured probability distribution of $P(\mathcal{O})$ for a given order parameter $\mathcal{O}$. We can visually inspect configurations to check the phase topology by first smearing the Polyakov loops at each spatial site. Three example bubble configurations are presented in Fig.~\ref{fig:vizs-three}, obtained via multicanonical run using $\langle l_{\theta} \rangle$ as the order parameter $\mathcal{O}$. 

\begin{figure}[h!]
   \centering 
   \includegraphics[width=0.85\textwidth]{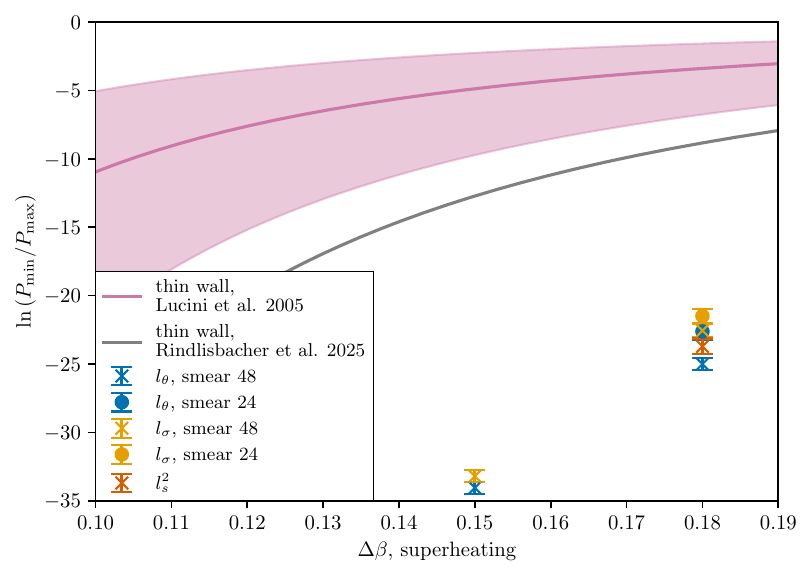}
   \caption{Critical bubble probability $\log\left[P(\mathcal{O}_{\mathrm{conf}})/P(\mathcal{O}_C)\right]$ obtained from the simulations on a $60^3 \times 6$ lattice, compared to the thin-wall estimate of the free energy, $-F_C/T$.  For $l_{\theta}$ and $l_{\sigma}$, two different levels of smearing are shown for $\beta_2$, and only one for $\beta_1$. The thin-wall estimate takes latent heat and surface tension measured from lattice as inputs, for the upper line from Ref.~\cite{Lucini:2005vg}. The error is from determination of these. Lower line is a preliminary result from Ref.~\cite{surfacetension}. }
\label{fig:result_free_energy}
\end{figure}

In Fig.~\ref{fig:lp-ltheta} the probability distribution measured with $\langle l_p \rangle$ as the order parameter of the multicanonical weight iteration is compared to the distribution obtained by using $\langle l_{\theta} \rangle$ instead. Marked on both histograms are the points where the configurations illustrated in Fig.~\ref{fig:vizs-three} fall, from left to right. The configuration shown in the middle is, according to the order parameter $\langle l_{\theta} \rangle$, near to the critical bubble. However, when using $\langle l_p \rangle$, the order parameter value is within the bulk phase peak. Most of the bubble branch is indeed hidden by the bulk peak, making the critical bubble completely unresolvable.

\subsection{Critical bubble free energy}

As described in section \ref{sect:nucleratecalc}, we measure the critical bubble free energy at two different levels of superheating, for one lattice size $N_s = 60, N_t = 6$, testing the three different smeared order parameters $\langle l_{s}^2 \rangle $, $\langle l_{\theta} \rangle $, $\langle l_{\sigma} \rangle $. The thermodynamic and continuum limits will be explored in an upcoming publication. 

\begin{figure}[h]
   \centering \includegraphics[width=0.5\textwidth]{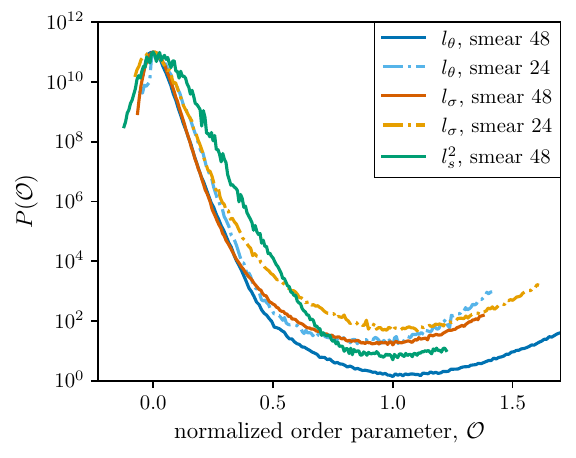} 
   \hspace{-3mm}
    \includegraphics[width=0.5\textwidth]{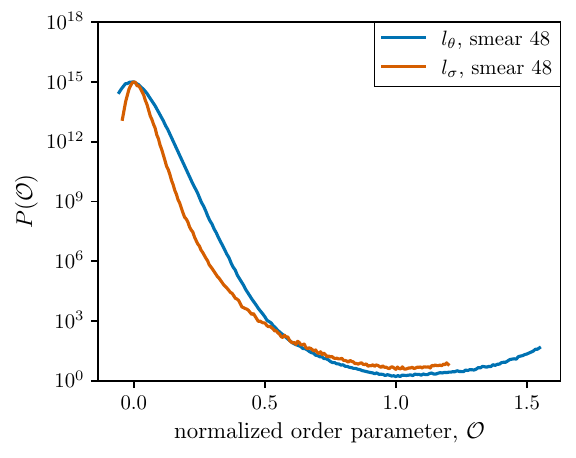}
   \caption{Qualitative comparison of histograms from different order parameters. Order parameters are normalized individually so that the minimum is at 1, the maximum at 0. \textbf{Left}: $\beta_2 = 44.742$. \textbf{Right}: $\beta_1 = 44.712$.}
   \label{fig:result_histos}
\end{figure}

In Fig.~\ref{fig:result_free_energy} the results for the free energy of the critical bubble are plotted together with the thin-wall estimate. The free energy is obtained by taking the minimum and maximum of the corresponding histogram (see Fig.~\ref{fig:result_histos}), and the errors via bootstrap.  The thin-wall estimate is most accurate for very small superheating, with our choices of $\beta_1 = 44.712$ and $\beta_2 = 44.742$ being superheated enough that we do not expect the thin-wall estimate to be very accurate. This is indeed what we see, with the difference in $F_C/T$ being from 10 to 25. This means the nucleation rate with the thin-wall would be an overestimate by a factor of $e^{10}$ to $e^{25}$, compared to the lattice result.

All of the order parameters had either 48 or 24 steps of smearing. Varying the level of smearing should have some effect on the critical bubble free energy. Some level of smearing is necessary to smooth the configurations enough to get rid of the majority of the UV fluctuations, but at some point more smearing is both computationally too expensive and physically unnecessary. Using 24 steps of smearing instead of 48, i.e. using half the number of steps, did have some effect on the free energy, but the results are within two standard deviations of each other. 

Interestingly, for this volume at $\beta_2$, $\langle l^2_s \rangle$ seems to be a good enough order parameter to resolve the bubble branch and the critical bubble, and gives comparable result to the other tested order parameters. However, quite a lot of smearing is required to make the critical bubble resolvable with $\langle l^2_s \rangle$, and by inspecting the histograms side by side, Fig.~\ref{fig:result_histos}, the bulk phase peak is wider than for $\langle l_{\theta} \rangle$ and $\langle l_{\sigma}\rangle$. We expect that at equal superheating but larger volume the bubble would not be resolvable with just $\langle l^2_s \rangle$, as the bulk phase peak becomes wider. 

\section{Conclusions}

We have resolved the critical bubble in the deconfinement phase transition of the SU(8) pure gauge model, and thus showed that this is possible in a confining theory. We have tested several different modified order parameters in order to not lose the signal under bulk phase fluctuations, and found clear improvement over standard Polyakov loop. 

The different order parameters do not agree within errors, but this is to be expected, as just the critical bubble free energy is not necessarily order parameter independent. The full bubble nucleation rate, which can be obtained in weakly coupled theories via evolving critical bubble configurations with real time simulations \cite{Moore:2000jw}, would in the end get rid of this effect, and the full rate would be order parameter independent. Unfortunately real time evolution methods are not available for strongly coupled SU($N$) theory.

Further investigation of the order parameters, and the infinite volume limit of the free energy of the critical bubble will be presented in an upcoming publication. Our result indicates that given some computational effort, for some confining theories the non-dynamical part of nucleation rate can indeed be estimated from the lattice with the multicanonical method.

\acknowledgments

The authors would like to thank CSC - IT Center for Science, Finland, for computational resources.
R.S. (ORCID ID 0000-0002-1461-2644) was supported by a working grant from the Magnus Ehrnrooth foundation and Research Council of Finland grant 349865. K.R. (ORCID ID 0000-0003-2266-4716) was supported by the Research Council of Finland grant 354572 and European Research Council grant 101142449. D.J.W. (ORCID ID 0000-0001-6986-0517) was supported by Research Council of Finland grants 324882, 349865 and 353131.

\bibliographystyle{JHEP}
\bibliography{references}

\providecommand{\href}[2]{#2}\begingroup\raggedright\begin{thebibliography}{10}

\bibitem{Huang:2020crf}
W.-C.~Huang, M.~Reichert, F.~Sannino and Z.-W.~Wang, \emph{{Testing the dark
  SU(N) Yang-Mills theory confined landscape: From the lattice to gravitational
  waves}}, \href{https://doi.org/10.1103/PhysRevD.104.035005}{\emph{Phys. Rev.
  D} {\bfseries 104} (2021) 035005}
  [\href{https://arxiv.org/abs/2012.11614}{{\ttfamily 2012.11614}}].

\bibitem{LatticeStrongDynamics:2020jwi}
{\scshape Lattice Strong Dynamics} collaboration, \emph{{Stealth dark matter
  confinement transition and gravitational waves}},
  \href{https://doi.org/10.1103/PhysRevD.103.014505}{\emph{Phys. Rev. D}
  {\bfseries 103} (2021) 014505}
  [\href{https://arxiv.org/abs/2006.16429}{{\ttfamily 2006.16429}}].

\bibitem{Bruno:2024dha}
M.~Bruno, N.~Forzano, M.~Panero and A.~Smecca, \emph{{Thermal evolution of dark
  matter in the early universe from a symplectic glueball model}},
  \href{https://arxiv.org/abs/2410.17122}{{\ttfamily 2410.17122}}.

\bibitem{Moore:2000jw}
G.D.~Moore and K.~Rummukainen, \emph{{Electroweak bubble nucleation,
  nonperturbatively}},
  \href{https://doi.org/10.1103/PhysRevD.63.045002}{\emph{Phys. Rev. D}
  {\bfseries 63} (2001) 045002}
  [\href{https://arxiv.org/abs/hep-ph/0009132}{{\ttfamily hep-ph/0009132}}].

\bibitem{Lucini:2005vg}
B.~Lucini, M.~Teper and U.~Wenger, \emph{{Properties of the deconfining phase
  transition in SU(N) gauge theories}},
  \href{https://doi.org/10.1088/1126-6708/2005/02/033}{\emph{JHEP} {\bfseries
  02} (2005) 033} [\href{https://arxiv.org/abs/hep-lat/0502003}{{\ttfamily
  hep-lat/0502003}}].

\bibitem{Lucini:2012wq}
B.~Lucini, A.~Rago and E.~Rinaldi, \emph{{SU($N_c$) gauge theories at
  deconfinement}},
  \href{https://doi.org/10.1016/j.physletb.2012.04.070}{\emph{Phys. Lett. B}
  {\bfseries 712} (2012) 279}
  [\href{https://arxiv.org/abs/1202.6684}{{\ttfamily 1202.6684}}].

\bibitem{surfacetension}
T.~Rindlisbacher, K.~Rummukainen and A.~Salami, \emph{{The confined-deconfined
  surface tension in SU(N) gauge theories at large N}}, {\emph{PoS} {\bfseries
  LATTICE2024} (2025) 406}.

\bibitem{Moore:2001vf}
G.D.~Moore, K.~Rummukainen and A.~Tranberg, \emph{{Nonperturbative computation
  of the bubble nucleation rate in the cubic anisotropy model}},
  \href{https://doi.org/10.1088/1126-6708/2001/04/017}{\emph{JHEP} {\bfseries
  04} (2001) 017} [\href{https://arxiv.org/abs/hep-lat/0103036}{{\ttfamily
  hep-lat/0103036}}].

\bibitem{Gould:2022ran}
O.~Gould, S.~G\"uyer and K.~Rummukainen, \emph{{First-order electroweak phase
  transitions: A nonperturbative update}},
  \href{https://doi.org/10.1103/PhysRevD.106.114507}{\emph{Phys. Rev. D}
  {\bfseries 106} (2022) 114507}
  [\href{https://arxiv.org/abs/2205.07238}{{\ttfamily 2205.07238}}].

\bibitem{Gould:2024chm}
O.~Gould, A.~Kormu and D.J.~Weir, \emph{{A nonperturbative test of nucleation
  calculations for strong phase transitions}},
  \href{https://arxiv.org/abs/2404.01876}{{\ttfamily 2404.01876}}.

\bibitem{Ferrenberg:1988yz}
A.M.~Ferrenberg and R.H.~Swendsen, \emph{{New Monte Carlo Technique for
  Studying Phase Transitions}},
  \href{https://doi.org/10.1103/PhysRevLett.61.2635}{\emph{Phys. Rev. Lett.}
  {\bfseries 61} (1988) 2635}.

\bibitem{Ferrenberg:1989ui}
A.M.~Ferrenberg and R.H.~Swendsen, \emph{{Optimized Monte Carlo analysis}},
  \href{https://doi.org/10.1103/PhysRevLett.63.1195}{\emph{Phys. Rev. Lett.}
  {\bfseries 63} (1989) 1195}.

\bibitem{Berg:1992qua}
B.A.~Berg and T.~Neuhaus, \emph{{Multicanonical ensemble: A New approach to
  simulate first order phase transitions}},
  \href{https://doi.org/10.1103/PhysRevLett.68.9}{\emph{Phys. Rev. Lett.}
  {\bfseries 68} (1992) 9}
  [\href{https://arxiv.org/abs/hep-lat/9202004}{{\ttfamily hep-lat/9202004}}].

\end{thebibliography}\endgroup

\end{document}